# CLASSIFYING THE GROUPS OF ORDER $pq$ IN LEAN

SCOTT HARPER AND PEIRAN WU

ABSTRACT. This note discusses our formalisation in Lean of the classification of the groups of order $pq$ for (not necessarily distinct) prime numbers $p$ and $q$, together with various intermediate results such as the characterisation of internal direct and semidirect products.

1. INTRODUCTION

The computer-assisted formalisation of mathematics is a process that sees proofs being transcribed into computer code in order to be rigorously checked by software. The validity of the resulting formal proofs rests entirely on the logical soundness of the software checker, virtually eliminating the risk of human error. The formal verification of the Feit–Thompson theorem [5] using the Coq theorem prover [1] drew great attention from the mathematical community, as did the recent Liquid Tensor Experiment [2], which formalised the proof of the Clausen–Scholze theorem in a programming language called Lean [3].

Lean has rapidly become the most actively used theorem prover in the last few years, with a powerful and ever-expanding mathematical library *mathlib* [10] that contains 1.5 million lines of code and boasts over 300 contributors. In a notable project enabled by mathlib and the Lean user community, the polynomial Freiman–Ruzsa conjecture in additive combinatorics saw its proof formalised within a month of preprint publication in 2023 [7, 4]. In contrast, many classical results in group theory remain missing from mathlib, and while there is little doubt of their mathematical correctness, their addition to mathlib is an essential step to formalising modern developments in the field. This paper contributes to the coverage of group theory in mathlib.

An influential theme in finite group theory is classification. The most spectacular example of a result in this direction is the Classification of Finite Simple Groups (CFSG). Needless to say, formalising the CFSG is far out of reach, but a second-generation proof that is self-contained modulo some explicitly outlined background references is nearing completion (ten of the predicted twelve volumes in this series have been published, beginning with [6]). However, this is just one of many classification results that have shaped contemporary group theory and others include the O'Nan–Scott Theorem (see [9], for example), which, roughly speaking, classifies the finite primitive permutation groups, modulo the CFSG. Formalising the O'Nan–Scott Theorem in Lean is the ongoing work of the second author.

One is also naturally inclined to classify the groups of a given order. All groups of a given prime order are cyclic and hence isomorphic to each other. This result has been formalised in Lean,[1] and it is the only such classification result in Lean to date, other than the classification of the groups of order four.[2]

Our main contribution is the following.

**Formalisation 1.** *The classification of the groups of order pq for prime numbers p and q.*

Note that in Formalisation 1 the prime numbers $p$ and $q$ need not be distinct.

Along the way, we formalise a number of other group theoretic results, which we expect to be of independent use in further formalisation. The most significant of these is the following (see Lemma 2.2 for the precise statement).

**Formalisation 2.** *The characterisation of internal direct and semidirect products.*

The formal statements of our main theorems are presented in Section 3 and the full Lean code is available at [8]. This code is being incorporated into mathlib.

---

[1] `isCyclic_of_prime_card` in `Mathlib.GroupTheory.SpecificGroups.Cyclic`

[2] `IsKleinFour.mulEquiv` in `Mathlib.GroupTheory.SpecificGroups.KleinFour`

In advanced undergraduate courses in group theory, classifying the groups of a particular order, or type of order, is often used to showcase the techniques in finite group theory that have been studied thus far and to motivate the study of further techniques. For instance, classifying the groups of prime order requires Lagrange's theorem, classifying the groups of order $p^2$ for a prime $p$ requires direct products and classifying the groups of order $pq$ for distinct primes $p$ and $q$ requires Sylow theory and semidirect products. Likewise, formalising the proofs of such results in Lean, gives a good indication of mathlib's coverage of basic group theory, and this is one motivation for the work in this paper. The intermediate results formalised in this project open up the way for further classification results to be formalised.

We conclude by commenting on the structure of this paper. We first discuss the underlying mathematics in Section 2 and then present the formal statements of our main results in Section 3. Section 4 concludes with a discussion of various considerations that we have to take into account while formalising these results, such as how finiteness is handled in mathlib, various implicit aspects of our results that need to be made explicit in the formalisation process and the fact that there are multiple natural ways to formulate a classification result.

2. Underlying mathematics

We begin with a natural language statement of the result we wish to formalise.

**Theorem 2.1.** *Let $p$ and $q$ be primes.*
  (a) *Every group of order $p^2$ is isomorphic to $C_{p^2}$ or the direct product $C_p \times C_p$.*
  (b) *Every group of order $pq$, with $p < q$, is isomorphic to $C_{pq}$ or, if $p \mid q-1$, the semidirect product $C_q \rtimes C_p$.*
  (c) *In particular, there exists a noncyclic group of order $pq$ if and only if $p = q$ or $p \mid q-1$ or $q \mid p-1$, in which case there is a unique noncyclic group of order $pq$ up to isomorphism.*

The expression "the semidirect product $C_q \rtimes C_p$" in part (b) is hiding various subtleties. More precisely, a semidirect product $C_q \rtimes_\varphi C_p$ depends on a choice of homomorphism $\varphi \colon C_q \to \mathrm{Aut}(C_p)$, but $\mathrm{Aut}(C_p) \cong C_{p-1}$, so there exists a nontrivial homomorphism $\varphi \colon C_q \to \mathrm{Aut}(C_p)$ if and only if $p \mid q-1$, and it is easy to prove that for all such nontrivial homomorphisms the semidirect products obtained are isomorphic. Therefore, "the semidirect product $C_q \rtimes C_p$" is shorthand for "a semidirect product $C_q \rtimes_\varphi C_p$ for any choice of nontrivial homomorphism $\varphi \colon C_q \to \mathrm{Aut}(C_p)$, noting that such a homomorphism exists and the group obtained is independent of the choice of $\varphi$, up to isomorphism". It is these familiar abuses of notation that have to be contended with when formalising statements of this sort, and we discuss this further in Section 4.

Before discussing how we formalise this theorem in Lean, let us briefly outline the familiar proof of this theorem. First note that part (c) is an immediate consequence of parts (a) and (b). To prove part (a), we let $G$ be a noncyclic group of order $p^2$, establish that $G$ is abelian, identify two distinct subgroups $P \leq G$ and $Q \leq G$ of order $p$, which are necessarily normal, and, via elementary arguments, show that $PQ = G$ and $P \cap Q = 1$ from which it follows that $G \cong P \times Q$. Similarly, to prove part (b), we let $G$ be a noncyclic group of order $pq$, identify subgroups $P \leq G$ of order $p$ and $Q \leq G$ of order $q$, via Sylow theory, show that $Q$ is a normal subgroup of $G$ and, via elementary arguments, show that $PQ = G$ and $P \cap Q = 1$, from which it follows that $G \cong Q \rtimes_\varphi P$ for some homomorphism $\varphi \colon P \to \mathrm{Aut}(Q)$. We then prove, as noted above, that a nontrivial homomorphism $\varphi \colon P \to \mathrm{Aut}(Q)$ exists if and only if $p \mid q-1$ and that, up to isomorphism, the semidirect product $Q \rtimes_\varphi P$ is independent of the choice of $\varphi$.

As this outline highlights, the characterisation of internal direct and semidirect products is a crucial ingredient in the proof of this theorem, and an important aspect of our project is to formalise these characterisations, which we state in natural language below.

**Lemma 2.2.** *Let $G$ be a group with subgroups $H$ and $N$ satisfying $HN = G$ and $H \cap N = 1$.*
  (a) *If $N$ is a normal subgroup of $G$, then there exists a homomorphism $\varphi \colon H \to \mathrm{Aut}(N)$ such that $G \cong N \rtimes_\varphi H$.*
  (b) *Moreover, if $H$ and $N$ are both normal subgroups of $G$, then $G \cong H \times N$.*

## 3. Main statements in Lean

This section is dedicated to presenting the main results that we formalise in Lean. For all Lean code given in this paper, certain imports from mathlib are required. Given the instability of mathlib's directory structure, we have not specified the exact imports here, but they can be found in the source code. We assume that the reader is familiar with the Lean language and the general aspects of the mathlib library, and we refer the reader to the Lean language reference[3] and the mathlib documentation[4] for further information. However, let us give an informal explanation of some aspects that are particularly relevant to our statements.

Group structures are formalised by the `Group` type class and the property of a group being cyclic is formalised by the `IsCyclic` type class. If `G` and `H` are groups, then `G →* H` and `G ≃* H`, roughly speaking, are the types of homomorphisms and isomorphisms from `G` to `H`, respectively, and `MulAut G` is defined as `G ≃* G`, which itself is an instance of a group. (Strictly speaking, these are the types of monoid homomorphisms and semigroup isomorphisms, but since any semigroup homomorphism between groups is a group homomorphism, this subtlety is not usually relevant in practice.)

If `G` is an instance of `Group`, then `Subgroup G` is the type of subgroups of `G`, which is an instance of the type class `CompleteLattice` of complete lattices. In particular, for subgroups `H` and `K` of a group `G`, one can refer to their meet `H ⊓ K` (which is their intersection) and their join `H ⊔ K` (which is the subgroup they generate), together with the least element `⊥` (which is the trivial group) and the greatest element `⊤` (which is the whole group).

Finiteness of a type is formalised by the `Finite` type class, and if `α` is a finite type, then `Nat.card α` is the cardinality of `α`. We define the new type `MulZMod n`, which is a concrete type of cardinality $n$ with a cyclic group structure. See Sections 4.1 and 4.2 for more detail.

We can now discuss the main theorems we formalise, which are presented in Figure 1.

```
variable {p q : ℕ} [hp : Fact p.Prime] [hq : Fact q.Prime]

theorem exists_card_eq_prime_mul_prime_and_not_isCyclic_iff :
    (∃ (G : Type) (_ : Group G), Nat.card G = p * q ∧ ¬IsCyclic G)
    ↔ p = q ∨ p ∣ q - 1 ∨ q ∣ p - 1 := ...

theorem nonempty_mulEquiv_of_card_eq_prime_mul_prime_of_not_isCyclic
    {G1 : Type*} [Group G1] (h1 : Nat.card G1 = p * q) (h1' : ¬IsCyclic G1)
    {G2 : Type*} [Group G2] (h2 : Nat.card G2 = p * q) (h2' : ¬IsCyclic G2) :
    Nonempty (G1 ≃* G2) := ...

theorem nonempty_mulEquiv_prod_of_card_eq_prime_pow_two_of_not_isCyclic
    {G : Type*} [Group G] (h : Nat.card G = p ^ 2) (h' : ¬IsCyclic G) :
    Nonempty (G ≃* MulZMod p × MulZMod p) := ...

theorem nonempty_mulEquiv_semidirectProduct_of_card_eq_prime_mul_prime
    (hpq : p < q) {G : Type*} [Group G] (h : Nat.card G = p * q) :
    ∃ φ : MulZMod p →* MulAut (MulZMod q),
    Nonempty (G ≃* MulZMod q ⋊[φ] MulZMod p) := ...

theorem nonempty_mulEquiv_semidirectProduct_of_card_eq_prime_mul_prime_of_
not_isCyclic
    (hpq : p < q) {G : Type*} [Group G] (h : Nat.card G = p * q)
    (h' : ¬IsCyclic G) (φ : MulZMod p →* MulAut (MulZMod q)) (hφ : φ ≠ 1 ) :
    Nonempty (G ≃* MulZMod q ⋊[φ] MulZMod p) := ...
```

Figure 1. Main theorems in Lean (without proof)

---

[3]https://lean-lang.org/doc/reference/latest/
[4]https://leanprover-community.github.io/mathlib4_docs/index.html

Comparing the formal statements in Figure 1 with the natural language statements in Theorem 2.1, we see that the first two theorems correspond to part (c), the next theorem corresponds to part (a) and the final two theorems correspond to part (b).

In the process of establishing these five main theorems, we formalise a number of intermediate results, including the characterisations of internal direct and semidirect products given in Lemma 2.2. These are given in Figure 2. We note that the result concerning direct products has a short proof using the result concerning semidirect products.

```
variable {G : Type*} [Group G]

noncomputable def mulEquivSemidirectProduct
    {N H : Subgroup G}  (nN : Subgroup.Normal N)
    (inf_eq_bot : N ⊓ H = ⊥) (sup_eq_top : N ⊔ H = ⊤)
    {φ : H →* MulAut N} (conj : φ = MulAut.conjNormal.restrict H):
    G ≃* N ⋊[φ] H := ...

noncomputable def mulEquivProd
    {N H : Subgroup G} (nN : Subgroup.Normal N) (nH : Subgroup.Normal H)
    (inf_eq_bot : N ⊓ H = ⊥) (sup_eq_top : N ⊔ H = ⊤) :
    G ≃* N × H := ...
```

Figure 2. Main lemmas on direct and semidirect products in Lean (without proof)

The statement in Figure 2 requires one further piece of explanation. With `G` a group, `H` a subgroup of `G` and `N` a normal subgroup of `G`, the homomorphism `G →* MulAut N` induced by conjugation is given by `MulAut.conjNormal` and the restriction `H →* MulAut N` is therefore given by `MulAut.conjNormal.restrict H`.

## 4. Implementation considerations

This final section discusses various considerations that have to be taken into account when formalising our main results.

### 4.1. Expressing a cyclic group of finite order.
It is easy to introduce a "cyclic group of order $n$" in Lean as follows.

```
variable {n : ℕ} [NeZero n] {G : Type*} [Group G] [IsCyclic G]
    (h : Nat.card G = n)
```

Translating this into natural language, we let $n$ be a natural number that is nonzero, let $G$ be a type that has a group structure and suppose that $G$ is cyclic and of cardinality $n$. The nonzero condition is essential as `Nat.card` is defined to be 0 for an infinite type (see Section 4.2 below).

However, it would not be as efficient to describe a group that is a semidirect product of two cyclic groups in a similar fashion. An alternative approach is to state that $G$ is isomorphic to a concrete group with an apparent structure.

Mathlib defines a concrete type `ZMod n` (given a natural number `n`) representing the integers modulo `n`. The following entries in mathlib state that `ZMod n` is a commutative ring, its additive group structure is cyclic and it is of cardinality `n`.

```
instance ZMod.commRing (n : ℕ) : CommRing (ZMod n)

instance ZMod.instIsAddCyclic (n : ℕ) : IsAddCyclic (ZMod n) := ...

theorem Nat.card_zmod (n : ℕ) : Nat.card (ZMod n) = n := ...
```

Here, `IsAddCyclic` is the type class for groups with the additive notation `+`, and is distinct from `IsCyclic`, which is the type class for groups with the multiplicative notation `*`.

To make the final three theorems in Figure 1 easier to state, we need a concrete multiplicative group. To this end, we define a new dependent type `MulZMod` as the multiplicative version of the additive group `ZMod`.

```
def MulZMod (n : ℕ) : Type := Multiplicative (ZMod n)
```

Put simply, `Mulplicative` converts a type with an additive structure to one with a multiplicative structure. It is coupled with `Additive` for the opposite conversion and both come with a suite of definitions and theorems that allow results to be transported between the corresponding structures, as seen in the following.

```
instance {n : ℕ} [NeZero n] : Group (MulZMod n) := Multiplicative.group

instance {n : ℕ} [NeZero n] : IsCyclic (MulZMod n) := isCyclic_multiplicative
```

The above also demonstrates type class inference in Lean. For example, in the proof of the second instance, we have not explicitly supplied the theorem that says `ZMod n` is an additive cyclic group, as Lean is able to find the instance `ZMod.instIsAddCyclic` automatically.

It would be annoying if one had to manually convert each result about multiplicative groups to one about additive groups. This is not necessary thanks to the `@[to_additive]` attribute, which, when applied a definition or theorem in the multiplicative theory, automatically generates the corresponding definition or theorem in the additive theory.

### 4.2. Expressing finiteness and cardinalities.

In addition to `Finite`, mathlib also defines a type class `Fintype` for expressing the finiteness of type. Given a type `α`, while `Finite α` is a `Prop`, `Fintype α` is not and carries an enumeration of all the elements of `α` as data. The practical differences between the two type classes are explained in the mathlib documentation under the definition of `Finite`.[5]

There are also two ways to express the cardinality of a type as a natural number, `Nat.card` and `Fintype.card`. The latter is only defined for types that are `Fintype`, while the former can be used on any type but evaluates to zero for infinite types. If `Nat.card α` is positive, then `Finite α` follows as a consequence.

By design, theorems in Lean should avoid `Fintype` wherever possible, which is a practice that we follow. There have been ongoing efforts to migrate old theorems that predate the introduction of `Finite` away from `Fintype`.[6] Notably, the signature of Lagrange's theorem was recently changed (see Figure 3). We have also contributed to the migration efforts as it is useful for avoiding `Fintype` in our theorems.

```
-- previously, Lagrange's theorem used `Fintype` and `Fintype.card`
theorem Subgroup.card_subgroup_dvd_card
    {α : Type u_1} [Group α] (s : Subgroup α) [Fintype α] [Fintype s] :
    Fintype.card ↥s ∣ Fintype.card α := ...

-- as of commit 480278c, it uses `Nat.card`
theorem Subgroup.card_subgroup_dvd_card
    {α : Type u_1} [Group α] (s : Subgroup α) :
    Nat.card ↥s ∣ Nat.card α := ...
```

FIGURE 3. Changes to the signature of Lagrange's theorem

---

[5]https://leanprover-community.github.io/mathlib4_docs/Mathlib/Data/Finite/Defs.html#Finite

[6]An example of work on this front is GitHub pull request #13431 on the mathlib4 repository: https://github.com/leanprover-community/mathlib4/pull/13431

### 4.3. Formulating a classification.

There are a number of ways one can state a classification result. For instance, when stating the classification of groups of order $p^2$ up to isomorphism in natural language, we begin with the hypothesis

> Let $G$ be a group of order $p^2$.

and then there are at least the following three options for the conclusion:

(a) Then $G$ is an elementary abelian $p$-group of exponent at most two.
(b) Then $G \cong \mathbb{Z}/p^2\mathbb{Z}$ or $G \cong \mathbb{Z}/p\mathbb{Z} \times \mathbb{Z}/p\mathbb{Z}$.
(c) Then if $G$ is not cyclic, we have $G \cong \mathbb{Z}/p\mathbb{Z} \times \mathbb{Z}/p\mathbb{Z}$.

The three options put emphasis on different aspects of the result. Option (a) gives a characterisation of the groups, while (b) and (c) give representatives of the isomorphism classes. In the latter case, we need to decide on appropriate representatives. As discussed in Section 4.1, we choose representatives expressed in terms of $\mathbb{Z}/n\mathbb{Z}$ under addition but written multiplicatively.

The difference between (b) and (c) lies in whether there is a more general hypothesis ("$G$ has order $p^2$" versus "$G$ has order $p^2$ and is not cyclic") or a more specific conclusion ("$G \cong \mathbb{Z}/p^2\mathbb{Z}$ or $G \cong \mathbb{Z}/p\mathbb{Z} \times \mathbb{Z}/p\mathbb{Z}$" versus "$G \cong \mathbb{Z}/p\mathbb{Z} \times \mathbb{Z}/p\mathbb{Z}$"). Option (b) demonstrates our knowledge of an exhaustive list of representatives and is useful for dividing a discussion into cases. (One could also point out that the representatives are themselves pairwise nonisomorphic.) Option (c), on the other hand, gives a sufficient condition for directly identifying the isomorphism class of a given group; this is particularly useful as the number of possibilities in the conclusion increases (for instance, in the classification of groups of order $p^3$).

Finally, we note that formalising a statement of the form $G \cong H$ involves a decision. We could give an explicit isomorphism by writing a `def` with return type `G ≃* H`, or we could merely assert that an isomorphism exists by writing a `lemma` or `theorem` with conclusion `Nonempty (G ≃* H)`. The former approach should be taken when it is worth keeping the exact definition of the isomorphism; for example, one may want to perform computations or prove properties of the isomorphism. In other cases, the latter is preferable, as proof irrelevance means the compiler does not need to load into its context any construction used to prove that the groups are isomorphic.

### 4.4. Formalising results on semidirect products.

The type `SemidirectProduct N H φ` represents the semidirect product of (multiplicative) groups `N` and `H` with respect to a homomorphism `φ` from `H` to the automorphism group of `N`.

After developing some useful lemmas about semidirect products (see Figure 4), we prove the key result `mulEquivSemidirectProduct`, shown in Figure 5, which corresponds to Lemma 2.2(a). In the spirit of stating definitions and theorems in as much generality as possible, we also prove an alternative version `mulEquivSemidirectProduct'`, which gives a sufficient condition for when a subgroup of a group is a semidirect product.

```
variable {N H : Type*} [Group N] [Group H]

lemma SemidirectProduct.card (φ : H →* MulAut N) :
    Nat.card (N ⋊[φ] H) = Nat.card N * Nat.card H := ...

def SemidirectProduct.mulEquivProd :
    N ⋊[1] H ≃* N × H where ...

variable {N₁ N₂ H₁ H₂ : Type*} [Group N₁] [Group N₂] [Group H₁] [Group H₂]

def SemidirectProduct.congr
    {φ₁ : H₁ →* MulAut N₁} {φ₂ : H₂ →* MulAut N₂}
    (f₁ : N₁ ≃* N₂) (f₂ : H₁ ≃* H₂)
    (h : ∀ n₁ : N₁, ∀ h₁ : H₁, φ₂ (f₂ h₁) (f₁ n₁) = f₁ ((φ₁ h₁) n₁) ) :
    N₁ ⋊[φ₁] H₁ ≃* N₂ ⋊[φ₂] H₂ where ...
```

Figure 4. Basic results on semidirect products

```
noncomputable def mulEquivSemidirectProduct
    {G : Type*} [Group G] {N H : Subgroup G} (h : Subgroup.Normal N)
    (inf_eq_bot : N ⊓ H = ⊥) (sup_eq_top : N ⊔ H = ⊤)
    {φ : H →* MulAut N} (conj : φ = MulAut.conjNormal.restrict H):
    G ≃* N ⋊[φ] H := ...

noncomputable def mulEquivSemidirectProduct'
    {G : Type*} [Group G]  {N H : Subgroup G} (h : Subgroup.Normal N)
    (inf_eq_bot : N ⊓ H = ⊥)
    {φ : H →* MulAut N} (conj : φ = MulAut.conjNormal.restrict H):
    (N ⊔ H : Subgroup G) ≃* N ⋊[φ] H := ...
```

FIGURE 5. Two versions of the characterisation of internal semidirect products

4.5. **Translating between additive groups and commutative rings.** A key ingredient in the proof of the main theorems is the fact that the automorphism group of a cyclic group of prime order $p$ is cyclic of order $p-1$.

We could prove this in Lean from first principles, but it is easier to use the following existing knowledge in mathlib about `ZMod n` as a commutative ring:

(i) `ZMod n` is a finite integral domain if `n` is prime.
(ii) `instIsCyclicUnitsOfFinite`: the unit group of a finite integral domain is cyclic.
(iii) `ZMod.card_units_eq_totient`: the number of units in `ZMod n` is equal to the Euler totient function of `n`.

Using this, we define an isomorphism from `AddAut (ZMod p)`, the automorphism group of `ZMod p` (as an additive group), to the unit group of `ZMod p` (as a commutative ring). We then transform it to an isomorphism from `MulAut (MulZMod p)` to the unit group of `ZMod p`. Finally, we use the above results to prove that `MulAut (MulZMod p)` is cyclic of order $p-1$. This is presented in Figure 6.

```
variable (p : ℕ) [Fact (p.Prime)]

def mulEquivAddAutZMod : AddAut (ZMod p) ≃* Units (ZMod p) where ...

def mulEquivMulAutMulZMod : MulAut (MulZMod p) ≃* Units (ZMod p) :=
    AddEquiv.toMultiplicative.mulEquiv.symm.trans <| addEquivAddAutZMod p

lemma mulAut_MulZMod_isCyclic : IsCyclic (MulAut (MulZMod p)) := ...

lemma card_mulAut_mulZMod :
    Nat.card (MulAut (MulZMod p)) = p - 1 := ...
```

FIGURE 6. Proving that $\mathrm{Aut}(C_p) \cong C_{p-1}$

**Acknowledgements.** The authors would like to thank the Lean community for their support during the project. The first author is an EPSRC Postdoctoral Fellow (EP/X011879/1). In order to meet institutional and research funder open access requirements, any Author Accepted Manuscript arising shall be open access under a Creative Commons Attribution (CC BY) licence with zero embargo.

Scott Harper, School of Mathematics and Statistics, University of St Andrews, St Andrews, KY16 9SS, UK

*Email address*: scott.harper@st-andrews.ac.uk

Peiran Wu, School of Mathematics and Statistics, University of St Andrews, St Andrews, KY16 9SS, UK

*Email address*: pw72@st-andrews.ac.uk